\documentclass[twocolumn,floats,floatfix,amssymb,nofootinbib,prd,superscriptaddress]{revtex4-1}
\usepackage{graphicx}
\usepackage{amsmath,amssymb}
\usepackage{amsfonts}
\usepackage{xspace} 
\usepackage[usenames]{color}
\usepackage{dcolumn}
\usepackage{bm}
\usepackage{mathrsfs}
\usepackage[colorlinks=true]{hyperref}
\usepackage[all]{hypcap} 
\usepackage[utf8]{inputenc} 

\usepackage{etoolbox}
\usepackage{tikz}
\def\be{\begin{equation}}
\def\ee{\end{equation}}
\def\bea{\begin{eqnarray}}
\def\eea{\end{eqnarray}}

\newcommand{\beq}{\begin{eqnarray}}
\newcommand{\eeq}{\end{eqnarray}}

\newcommand{\tn}{\textnormal}

\begin{document}

\title{Gravitational waves in massive gravity theories: \\
waveforms, fluxes and constraints from extreme-mass-ratio mergers}

\author{Vitor Cardoso}\email{vitor.cardoso@tecnico.ulisboa.pt}
\affiliation{Centro de Astrof\'{\i}sica e Gravita\c c\~ao  - CENTRA, Departamento de F\'{\i}sica, Instituto Superior T\'ecnico - IST, Universidade de Lisboa - UL, Av. Rovisco Pais 1, 1049-001 Lisboa, Portugal}
\affiliation{Theoretical Physics Department, CERN 1 Esplanade des Particules, Geneva 23, CH-1211, Switzerland}
\author{Gon\c calo Castro}\email{gcabritac@gmail.com}
\affiliation{Centro de Astrof\'{\i}sica e Gravita\c c\~ao  - CENTRA, Departamento de F\'{\i}sica, Instituto Superior T\'ecnico - IST, Universidade de Lisboa - UL, Av. Rovisco Pais 1, 1049-001 Lisboa, Portugal}
\author{Andrea Maselli}\email{andrea.maselli@tecnico.ulisboa.pt}
\affiliation{Centro de Astrof\'{\i}sica e Gravita\c c\~ao  - CENTRA, Departamento de F\'{\i}sica, Instituto Superior T\'ecnico - IST, Universidade de Lisboa - UL, Av. Rovisco Pais 1, 1049-001 Lisboa, Portugal}
\affiliation{Dipartimento di Fisica, ``Sapienza'' Universit\`a di Roma, Piazzale 
Aldo Moro 5, 00185, Roma, Italy}

\begin{abstract}
Is the graviton massless? This problem was addressed in the literature at a phenomenological level, using modified dispersion relations for 
gravitational waves, in linearized calculations around flat space. Here, we perform a detailed analysis of the gravitational waveform
produced when a small particle plunges or inspirals into a large non-spinning black hole. Our results should presumably also describe the gravitational collapse to black holes and explosive events such as supernovae. In the context of a theory with massive gravitons and screening, merging objects up to $1\,{\rm Gpc}$ away or collapsing stars in the nearby galaxy may be used to constrain the mass of the graviton to be smaller than $\sim 10^{-23}\,{\rm eV}$, with low-frequency detectors. Our results suggest that the absence of dipolar gravitational waves from 
black hole binaries may be used to rule out entirely such theories.
\end{abstract}

\date{\today}
\maketitle 

%%%%%%%%%%%%%%%%%%%%%%%%%%%%%%%%%%%%%%%%%%%%%%%%%
\section{Introduction}
%%%%%%%%%%%%%%%%%%%%%%%%%%%%%%%%%%%%%%%%%%%%%%%%%
General Relativity (GR) is special and unique, in a very precise mathematical sense~\cite{Berti:2015itd,Barack:2018yly}.
Nevertheless, several arguments suggest that such an elegant theory cannot easily accommodate neither the ultraviolet nor the infrared description of the universe. Simultaneously, observations of large-scale phenomena indicate that either the matter sector or the gravitational interaction require a better understanding. In other words, extensions of GR are welcome. One of the possible extensions draws inspiration from the standard model of particle physics, and consists in
allowing for a massive graviton~\cite{Hinterbichler:2011tt,deRham:2014zqa,Barack:2018yly,Hassan:2012ka}.

Bounds on such theories can be imposed via gravitational-wave (GW) emission and propagation mechanisms. These include:

\noindent i. Modified dispersion relations for GWs, {\it assuming} that their generation is as in GR~\cite{Will:1997bb,TheLIGOScientific:2016src,Abbott:2017oio}.

\noindent ii. The spin-down of black holes (BHs), caused by superradiant instabilities~\cite{Brito:2013wya}.

\noindent iii. Changes in the orbital period of binary pulsars, caused by a different energy flux~\cite{Finn:2001qi}.

Other mechanisms may also help in bounding the graviton mass, such as modifications of the GW memory effect~\cite{Kilicarslan:2018bia}.
There are no constraints using directly the measured properties of GWs, without any assumption on the production
mechanism. Our main concern here is precisely to compute the gravitational waveform and fluxes from the merger of two compact objects,
using the strong-field regime of massive gravity theories. We consider the ghost-free theory describing two interacting spin-2 fields
described in detail in the supplementary material.

Following all observational evidence thus far, we consider only BHs which are as similar as possible to those in GR; in particular, we study Schwarzschild BHs which are also exact solutions of massive bi-gravity theories. We focus on the truly unique features of massive gravity theories: the extra polarizations with respect to GR and their signatures on the GW emission.
We thus consider mergers of extreme-mass ratio objects in which the massive one is a Schwarzschild BH. We will show that
the extra degrees of freedom give rise to substantially different GW-signals, even when the underlying backgrounds are exactly the same.

The calculations can, in principle, encompass also collapsing objects as long as the final state is a BH. 
Finally, the extrapolation to nearly-equal mass objects 
allows us to infer physics of interest to Earth-based detectors.

Throughout this work we use geometrized units, in which $G=c=1$.

%%%%%%%%%%%%%%%%%%%%%%%%%%%%%%%%%%%%%%%%%%%%%%%%%%%%%%%%%%%%%%%%%%%
\section{Formalism and master equations}\label{Sec:metricequations}
%%%%%%%%%%%%%%%%%%%%%%%%%%%%%%%%%%%%%%%%%%%%%%%%%%%%%%%%%%%%%%%%%%%

In our framework, a small point particle is orbiting, or merging with, a massive Schwarzschild BH of mass $M$. This system may 
model the merger of a neutron star with a stellar-mass or a supermassive BH, but it may well describe qualitatively the merger of 
two equal-mass BHs as well. In fact, the lesson from the two-body problem in GR is that perturbation 
theory is able to account for this process even at a quantitative level~\cite{Cardoso:2014uka}. The point particle moves 
on a spacetime geodesic $y_p^{\mu}(\tau)=(t_p(\tau),r_p(\tau),\theta_p(\tau),\varphi_p(\tau))$, with $\tau$ being the test body 
proper time. The particle is taken to be pointlike and described by the stress-energy tensor
\begin{equation}
T^{\mu\nu}=m_p\int (-g)^{1/2} u^\mu u^\nu \delta^{(4)}(x^\beta-y_p^\beta)d\tau\ ,\label{stressnergy}
\end{equation}
where $m_p$ is the rest mass of the test particle and $u^\mu=dy_p^{\mu}/d\tau$ its 4-velocity. The point particle stress
slightly disturbs the background geometry $\bar{g}_{\mu\nu},\, \bar{f}_{\mu\nu}$ (the theory has two metrics) describing 
the BH and a graviton of mass $\mu$. The latter is given by a specific combination of the coupling parameters of the 
theory~\cite{Brito:2013wya}. Here, we study backgrounds for which the two  
metrics $\bar{g}_{\mu\nu}$ and $\bar{f}_{\mu\nu}$ are proportional, leading to geometries which coincide with those of GR~\cite{Hassan:2012wr}. The stress energy-tensor contributes with fluctuations $(\delta g_{\mu\nu}, \delta f_{\mu\nu})$, which we analyse in tensor spherical harmonics and Fourier decompose. Details are left for the the supplementary material.

%%%%%%%%%%%%%%%%%%%%%%%%%%%%%%%%%%%%%%%%%%%%%%%%%%%%%%%%%%%%%%%%%%%%%%
\subsection{Head-on collisions} \label{sec:headon_eqs}
%%%%%%%%%%%%%%%%%%%%%%%%%%%%%%%%%%%%%%%%%%%%%%%%%%%%%%%%%%%%%%%%%%%%%%

Hereafter, we consider two prototypical dynamical processes: radial infall corresponding to head-on collisions, and pure equatorial 
motion corresponding to quasicircular inspirals (once radiation reaction is taken into account). The complete expressions for the 
source components in these two specific configurations are shown in Sec.~IV of the supplementary material. 
For radial motion, axial perturbations are not excited. The multipolar expansion then describes only polar-type perturbations with $
\ell\geq0$. Of these, the $\ell\geq 2$ equations contain small $\mu$-dependent corrections to the GR expressions.
We do not consider these any further~\footnote{Such corrections were studied in some detail in the weak-field, slow-motion limit elsewhere~\cite{Finn:2001qi}.} and focus on the truly unique properties of massive gravity: the presence of new degrees of freedom, described by the $\ell=0$ and $\ell=1$ modes.

For the monopole, $\ell=0$ mode, the number of perturbation functions reduces to the four metric components $(H_0,H_1,H_2,K)$ 
(see supplementary material and Ref.~\cite{Brito:2013wya}). Through the following transformation:
\begin{equation}
K=\frac{\sqrt{-4 \mu ^2 M+\mu ^4 r^3+2 \mu ^2 r+4 r \omega ^2}}{r^{5/2}}\varphi_0\ ,
\end{equation}
we obtain a single wave equation for $\varphi_0$:
\begin{equation}
\frac{d^2\varphi_0}{dr_{\star}^2} + [\omega^2-V^{\ell=0}_\tn{pol}(r,\omega)]\varphi_0 = {\bf {\cal S}}^{\ell=0}_\tn{pol}\ .\label{eq:l0polar}
\end{equation}
Here, $V^{\ell=0}_\tn{pol}(r,\omega)$ is a radial potential whose expression is lengthy and not very illuminating, while $r_\star$ is a tortoise coordinate defined by $dr_\star/dr=1/f$. The potential 
$V^{\ell=0}_\tn{pol}(r,\omega)\sim \mu^2$ at large spatial distances, and it vanishes close to the BH horizon. The source 
term $ {\bf {\cal S}}^{\ell=0}_\tn{pol}$ depends on the radial position and on the energy with which the point particle is colliding.
In the highly relativistic regime,
\begin{equation}
{\bf {\cal S}}^{\ell=0}_\tn{pol}=\frac{8 \sqrt{2} m_p\gamma (r-2 M) \left(\mu ^2 r+2 i \omega \right) e^{i \omega  t_p(r)}}{\sqrt{r} \left(-4 \mu ^2 M+\mu ^4 r^3+2 \mu ^2 r+4 r \omega ^2\right)^{3/2}}\ ,\label{eq:l0polar_source}
\end{equation}
where $\gamma$ is the Lorentz boost factor of the test particle at large spatial separations. Note that the $z$-axis is chosen to coincide with the particle trajectory, hence only $m=0$ modes are excited.

For the dipole $\ell=1$ term the perturbations are completely determined by two coupled equations for $K$ and $\eta_1$, which 
can be recast in a linear form as:
\begin{equation}
\left[\frac{d}{dr_\star}+V^{\ell=1}_\tn{pol}(r)\right]{\bf \Sigma}={\bf {\cal S}}^{\ell=1}_\tn{pol} \ , \label{poleql1}
\end{equation}
where ${\bf \Sigma}=(K,\eta_1,dK/dr_\star,d\eta_1/dr_\star)^\tn{T}$, and $V^{(1)}_\tn{pol}$ is a $4\times 4$ 
matrix which is shown within the supplementary material. 
For a radial infalling particle with a relativistic boost factor, the source vector is simply given by 
\be
{\bf {\cal S}}^{\ell=1}_\tn{pol}=(0,0,S_K,S_{\eta_1})=(0,0,f(r)/r,1)S_{\eta_1}\ ,\label{sourcel10radial}
\ee
where $f(r)=(1-2M/r)$ and
\be
S_{\eta_1}=-\frac{8\sqrt{6}m_p\gamma(2+r^2\mu^2+2i r\omega)e^{i\omega t_p(r)}}{4M r^2\mu^2-8M-6r^3\mu^2-r^5\mu^4-4r^3\omega^2
}\ .\label{sourcel1radial}
\ee

%%%%%%%%%%%%%%%%%%%%%%%%%%%%%%%%%%%%%%%%%%%%%%%%%%%%%%%%%%%%%%%%%%%%%%
\subsection{Quasi-circular inspirals} \label{sec:inspiral_eqs}
%%%%%%%%%%%%%%%%%%%%%%%%%%%%%%%%%%%%%%%%%%%%%%%%%%%%%%%%%%%%%%%%%%%%%%

For circular motion, the only non-trivial new degree of freedom is the dipolar-polar component.
Our system of equations can be written as
\beq
&&K''+ a_1 K'+a_2K+a_3\eta_1'+a_4\eta_1=S_1\delta(r-r_p)\,,\label{eq:K}\\
&&\eta_1''+ b_1 \eta_1'+b_2\eta_1+b_3K'+b_4K=S_2\delta(r-r_p)\,,
\eeq
where primes stand for tortoise derivatives, and $r_p$ is the orbital radius of the test particle.
The system above can be cast in the form
\begin{equation}
\left[\frac{d}{dr_\star}+V^{\ell=1}_\tn{pol}(r)\right]{\bf \Sigma}={\bf {\cal S}}^{\ell=1}_\tn{circ} \,, \label{l1poleq}
\end{equation}
being ${\bf \Sigma}=(K,\eta_1,dK/dr_\star,d\eta_1/dr_\star)^\tn{T}$ and ${\bf {\cal S}}^{\ell=1}_\tn{circ}=(0,0,S^\tn{circ}_K,S^\tn{circ}_{\eta_1})$. 
We solve eq.~\eqref{l1poleq} by first constructing a $4\times 4$ fundamental matrix $X$ built with the homogenous solutions of 
the previous system (see Sec.~V of the supplementary), which yields the general solution: 
\begin{equation}
{\bf \Sigma}(\omega,r)=X\int^{\infty}_{-\infty} X^{-1}{\bf {\cal S}}^{\ell=1}_\tn{circ} dr_\star\ .\label{circsol}
\end{equation}
Note that the source vector contains a linear combinations of the Dirac delta and its first derivative. 
Therefore, integrating by part eq.~\eqref{circsol} we can immediately obtain an explicit form for the metric 
functions, ${\bf \Sigma}(\omega,r)=X[{\bf A}+{\bf B}]$, where ${\bf A}$ and ${\bf B}$ are two 
vectors given by: 
\begin{subequations}
\begin{align}
{\bf A}=&\left(1-\frac{2M}{r_p}\right)X^{-1}(r_p){\bf {\cal S}}^{\ell=1}_\tn{circ}(r_p)\ ,\\
{\bf B}=&-\frac{d}{dr}\left[\left(1-\frac{2M}{r}\right)X^{-1}{\bf {\cal S}}^{\ell=1}_\tn{circ}\right]_{r=r_p}\ .
\end{align}
\end{subequations}
%
%%%%%%%%%%%%%%%%%%%%%%%%%%%%%%%%%%%%%%%%%%%%%%%%%
\section{Numerical results}\label{sec:numerical}
%%%%%%%%%%%%%%%%%%%%%%%%%%%%%%%%%%%%%%%%%%%%%%%%%

In this section we describe the numerical results obtained by solving the systems of ODEs for the 
monopole and dipole component of the polar sector. As described in Sec.~\ref{Sec:metricequations}, 
we consider circular and radial trajectories: for both the configurations, axial modes are 
not excited, as the source terms vanish. We integrate eq.~\eqref{eq:l0polar} and eqns.~\eqref{poleql1} 
through a Green function approach, with appropriate boundary conditions at the BH horizon and at spatial 
infinity (see the supplementary material for further details).

%%%%%%%%%%%%%%%%%%%%%%%%%%%%%%%%%%%%%%%%%%%%%%%%%%%%%%%%%%%
\subsection{Head-on collisions}\label{sec:numerical_headon}
%%%%%%%%%%%%%%%%%%%%%%%%%%%%%%%%%%%%%%%%%%%%%%%%%%%%%%%%%%%
%
\begin{figure}[th]
\centering
\includegraphics[width=5.5cm]{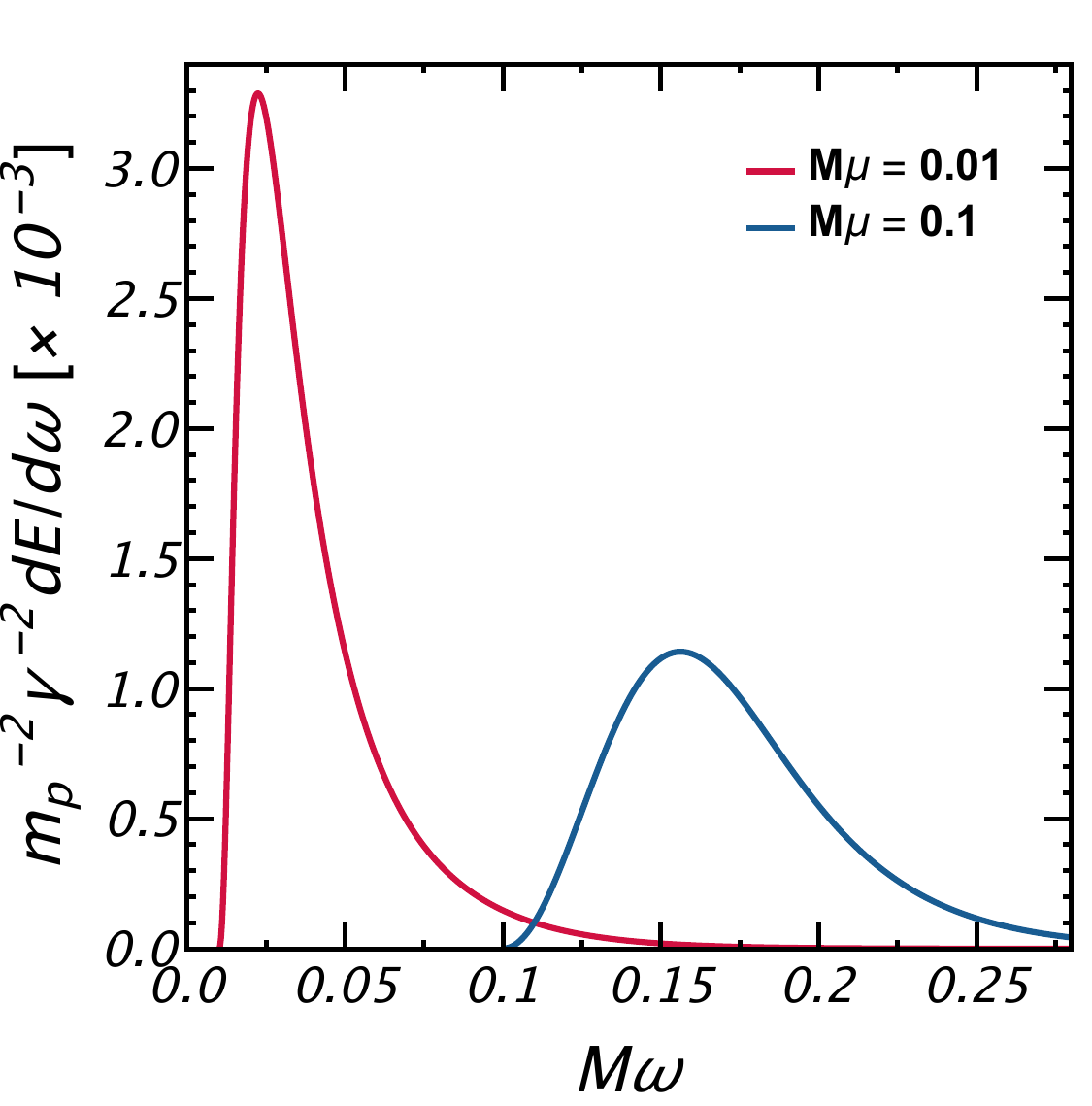}
\caption{GW energy spectrum $dE/d\omega$ for the $\ell=0$ polar mode, with $M\mu=(0.1,0.01)$ 
and a radial infalling particle.} 
\label{fig:l0_amp} 
\end{figure}
\begin{figure}[th]
\begin{tabular}{c}
\includegraphics[width=6cm]{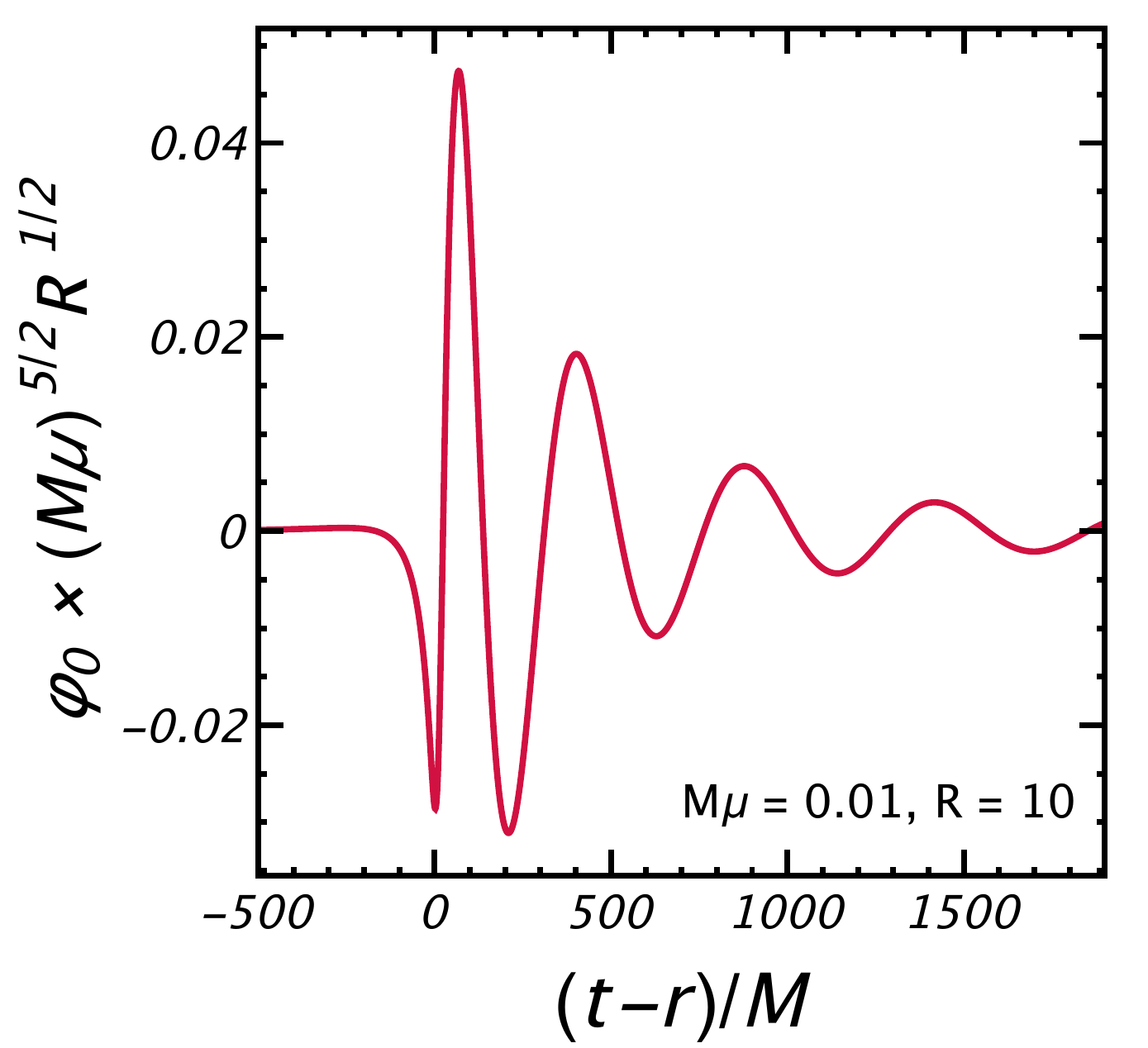}\\
\includegraphics[width=6cm]{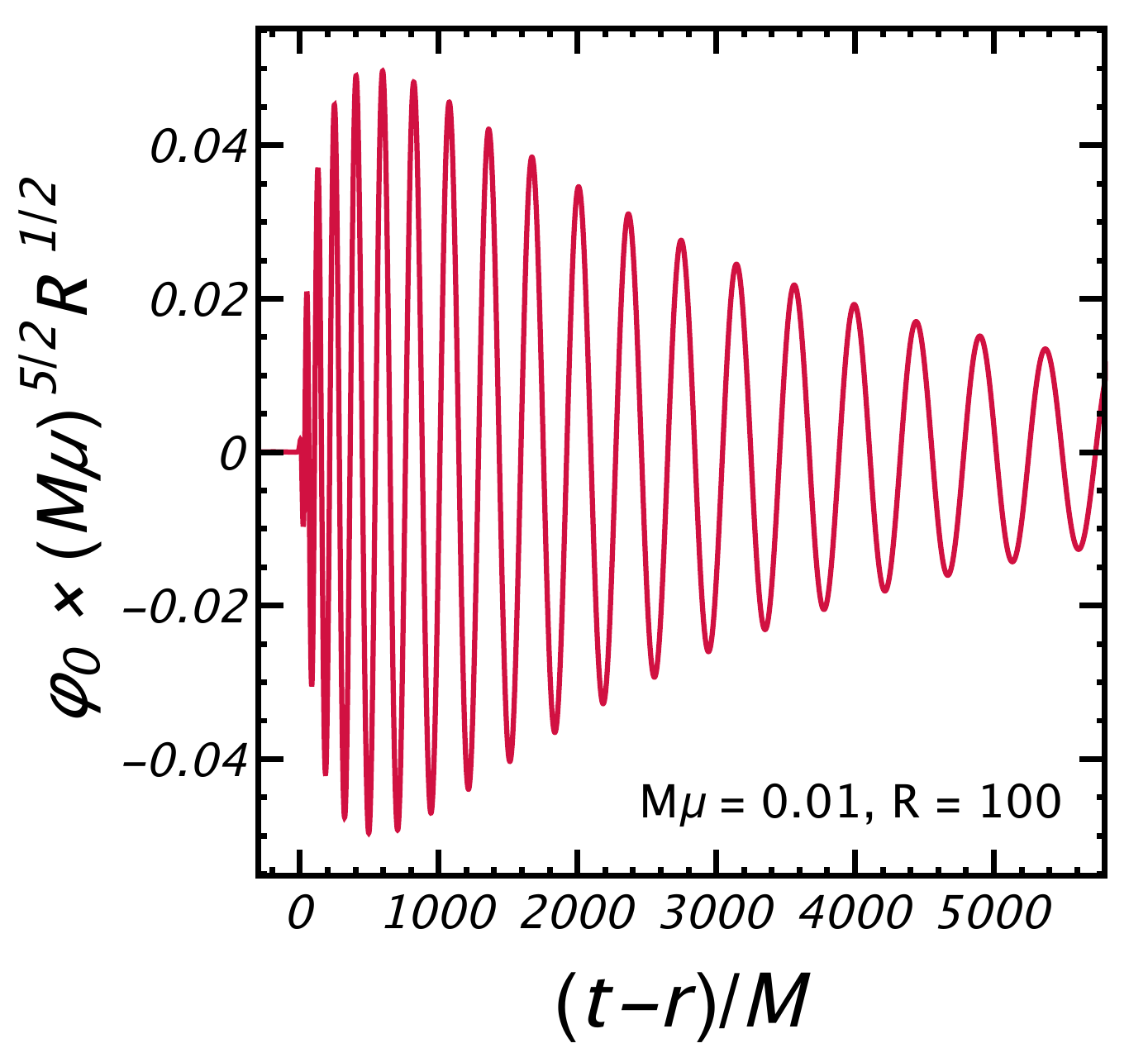}
\end{tabular}
\caption{Gravitational waveforms for the $\ell=0$ component of the polar sector, and a radial infalling particle, as a function of the retarded time $(t-r)/M$. 
We consider $M\mu=0.01$ and different extraction radius at $R=(10,100)$. The waveform scales trivially with the BH mass $M$ and the particle mass and boost $m_p, \gamma$, according to eq.~\eqref{peak_time}. \label{fig:l0_polar} } 
\end{figure}

For head-on collisions, the waveform amplitude scales linearly with the mass of the infalling point particle, and the only free parameter is the relative velocity at large distances. We fix this to be relativistic, and we find, as expected, that the amplitude then scales linearly with the boost factor $\gamma$. Although our formalism includes the general case, relativistic collisions should mimic well the late stages of inspiral. In addition, and perhaps more important for us here, they should also describe even explosive events such as supernovae. In theories of massive gravity, even spherically symmetric explosive events release a non-negligible amount of radiation in the monopole mode.

The energy spectrum $dE/d\omega$ for the monopole perturbation is shown 
in Fig.~\ref{fig:l0_amp} as a function of the frequency $\omega$, for a head-on collision.
The spectrum peaks close to the value of the graviton mass, and quickly decays to zero for higher frequencies. The total integrated energy is not shown but it scales like $E_{\rm tot}\sim 0.01 \mu m_p^2\gamma^2$ at small couplings $M\mu$.

Knowing the solution in the frequency domain, we can immediately compute the GW signal as a function of the retarded time 
by simply applying a Fourier transform to $\varphi_0(\omega)$. This is shown in Fig.~\ref{fig:l0_polar} for two values 
of the extraction radii $R=r\mu=(10,100)$~\cite{extraction_radii}. It is important to highlight that GWs in theories of massive gravity are {\it dispersed}: the waveform at large distances is no longer a function only of $t-r$. This property is apparent in Fig.~\ref{fig:l0_polar} and was also recently discussed in other setups~\cite{Sperhake:2017itk}. We find that the peak of the (time-domain) waveform can be described by the following scaling,
\be
\varphi_0^{\rm peak}\sim  \kappa\frac{m_p\gamma M^2}{(M\mu)^{5/2}R^{1/2}}\,,\label{scalingl0}
\ee 
where $\kappa\simeq0.055$, when the extraction radius $R>1$. This is not too surprising, given the $\mu$-dependence of the source term,
eq.~\eqref{sourcel1radial}, at low frequencies $\omega\sim \mu$. Our results indicate that the peak of the amplitude, with respect to the beginning of the signal, can be approximated by the following law,
\beq
&&(t-r)^{\rm peak}\sim (M\mu)^{-0.34} MR\nonumber\\
&\sim& 1800 \left(\frac{M}{M_{\odot}}\right)^{0.66}\left(\frac{\mu}{10^{-23}{\rm eV}}\right)^{0.66}\frac{r}{8{\rm Kpc}}{\rm secs}\,.\nonumber
\eeq

When expressed in terms of physical metric perturbations, we find
\beq
K^{\rm peak}&=&\kappa\frac{m_p}{M}\left(\frac{M}{r}\right)^{3/2}\frac{1}{M\mu} \label{peak_time}\\
&\sim&10^{-16}\frac{m_p\gamma}{0.01M}\sqrt{\frac{M}{M_{\odot}}}\left(\frac{8\,{\rm Kpc}}{r}\right)^{3/2}\frac{10^{-23}\,{\rm eV}}{\mu}\nonumber\\
&\sim&10^{-22}\frac{m_p\gamma}{0.01M}\sqrt{\frac{M}{M_{\odot}}}\left(\frac{{\rm Gpc}}{r}\right)^{3/2}\frac{10^{-25}\,{\rm eV}}{\mu}\ .\nonumber
\eeq
These numbers are encouraging, however the large-amplitude signals carry a low-frequency content $\omega \sim \mu$, corresponding to a frequency~\cite{Brito:2015yfh}
\be
f\sim 2.5\times 10^{-9}\left(\frac{\mu}{10^{-23}\,{\rm eV}}\right)\,{\rm Hz}\,.
\ee
Thus, observations of these signals will require low-frequency sensitive detectors.

At late times and large extraction radii, the waveform is exponentially damped. We cannot rule out power-law decay at very late times.
We have searched for the characteristic ringdown modes in this theory and find both good agreement with previously reported values~\cite{Brito:2013wya} and with the ones inferred from the time-domain waveforms. We note in particular the presence of an unstable mode, which does not seem to be significantly excited - on these timescales.
\begin{figure}[th]
\includegraphics[width=6cm]{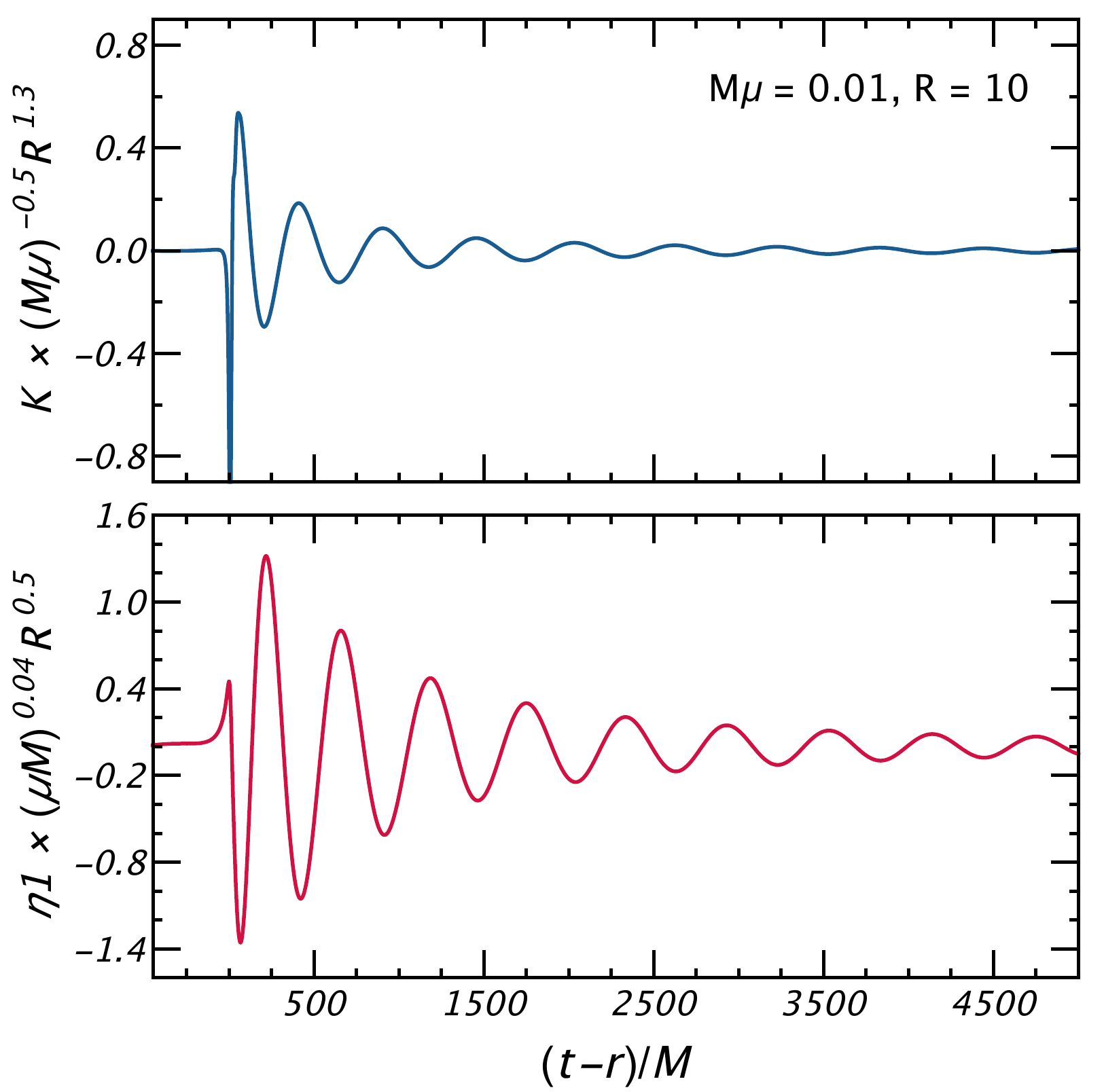}
\caption{Waveforms obtained for the $\ell=1$ polar mode, derived for a radial infalling particle 
with source term given by eq.~\eqref{sourcel1radial}, as a function of the retarded time $(t-r)/M$. 
The panel refers to $M\mu=0.01$ at extraction radii $R\mu=10$. The overall behavior is the same as the 
monopole $\ell=0$ mode.
The maximum amplitudes of the two metric functions scale as $K^{\rm peak}\sim m_p\gamma \delta_1\sqrt{M\mu}/R^{1.3}$ and 
$\eta_1^{\rm peak}\sim m_p\gamma M\delta_2(M\mu)^{-0.04}/R^{1/2}$
where $(\delta_1,\delta_2)\simeq(0.84,1.3)$ (these expressions also provide the scaling with $M, m_p, \gamma$).} 
\label{fig:l1_radial_polar} 
\end{figure}
Waveforms for the $\ell=1$ mode are shown in Fig.~\ref{fig:l1_radial_polar} (again for relativistic collisions). 
The maximum value of the amplitudes can be described again with a scaling factor of the form given by 
eq.~\eqref{scalingl0} (see caption of Fig.~\ref{fig:l1_radial_polar}).

Our results can also be applied to spherically symmetric collapse: in such a case, the source term is trivially replaced
by a spherically-symmetric shell; the final source term is unchanged. Even if $1\%$ of the star's rest mass is involved in the collapse, our results indicate that the peak waveform is detectable when $\mu$ is small enough. In fact, eq.~\eqref{peak_time} implies that stronger constraints can be obtained via (non-) observations of GWs from collapsing stars in our galaxy. Such conclusions are consistent also with recent results of core-collapse supernovae in massive scalar-tensor theories of gravity~\cite{Sperhake:2017itk}.

%%%%%%%%%%%%%%%%%%%%%%%%%%%%%%%%%%%%%%%%%%%%%%%%%%%%%%%%%%%%%%%%%%%%%%%
\subsection{Particles in circular motion}\label{sec:numerical_circular}
%%%%%%%%%%%%%%%%%%%%%%%%%%%%%%%%%%%%%%%%%%%%%%%%%%%%%%%%%%%%%%%%%%%%%%%
%
\begin{figure}[th]
\includegraphics[width=6.6cm]{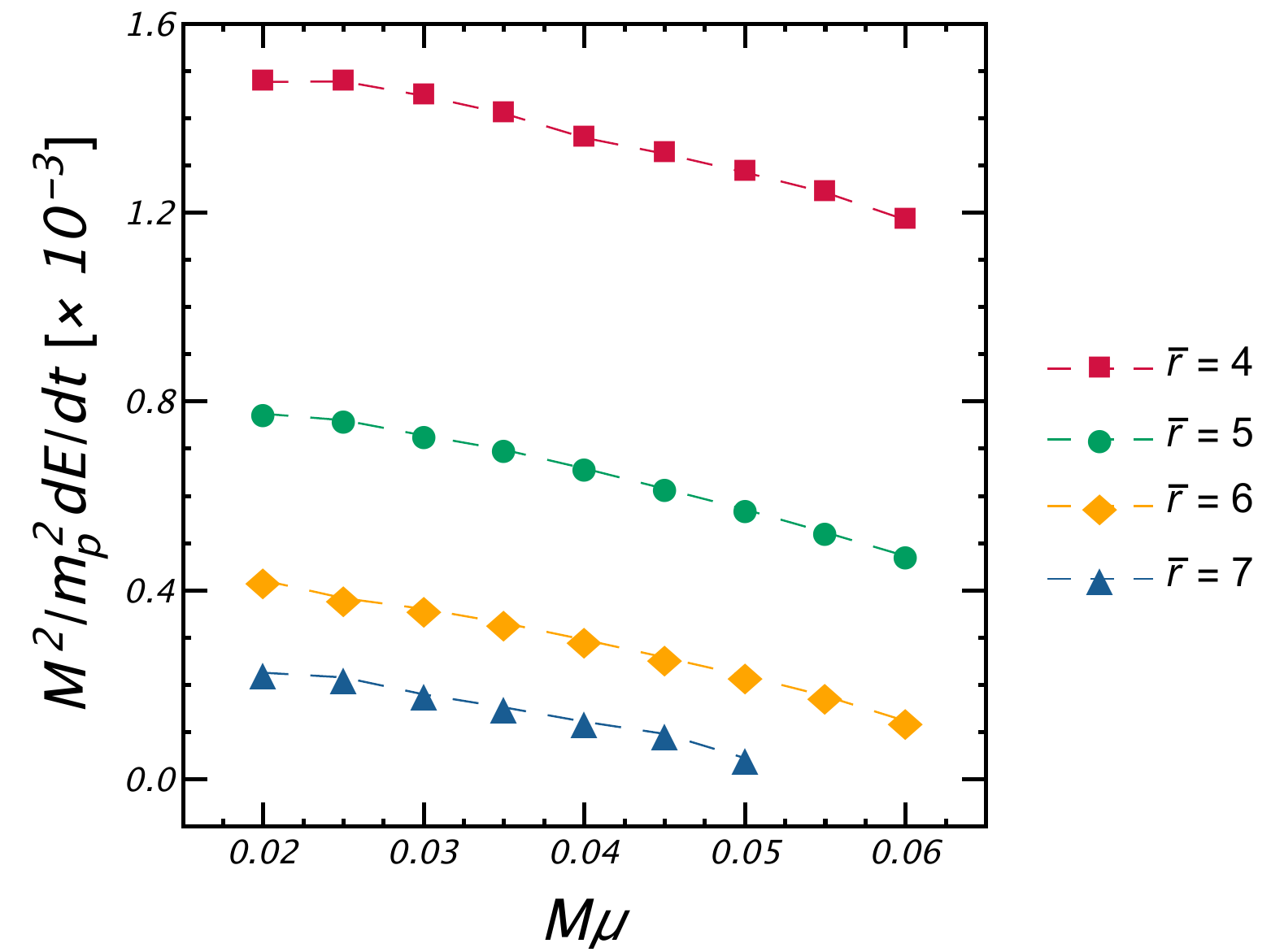}
\caption{GW luminosity $dE/dt$ for the $(\ell,m )=(1,1)$ polar mode as a function of the 2-spin field mass $M\mu$ 
for different radius $\bar{r}=r_p/M$ of the test particle on circular orbits around the BH.} 
\label{fig:l1_circular_luminosity} 
\end{figure}
Quasi-circular inspirals in the weak-field, slow-motion approximation have been used to impose constraints on massive theories of gravity
using pulsar timing observations~\cite{Finn:2001qi}. Those constraints used only corrections -- which scaled like $\mu^2$ -- to the quadrupole formula. Our results include relativistic motion in strong-gravity situations. In GR, particles in circular motion excite only quadrupolar or higher modes. As we saw, a new, dipolar mode arises in massive gravity, the energy flux of which is shown in Fig.~\ref{fig:l1_circular_luminosity}.

For a particle in a circular orbit of radius $r_p$, our results indicate that the flux in the $\ell=m=1$ mode scales like $1/r_p^4$, so truly a dipolar behavior, with BHs having a nontrivial dipolar charge in this theory. Furthermore, the charge is non-negligible at small $M\mu$. We find a flux $dE/dt\sim 0.6 m_p^2M^2/r_p^4$~\cite{vainshtein}. On the other hand, the quadrupole formula in GR predicts a quadrupolar emission $dE/dt=(32/5)m_p^2M^3/r_p^5$. This is one of our main results: the dipolar emission in massive gravity theories dominates the GR quadrupolar term, at arbitrarily small $\mu$. Thus, observations of binary BHs can potentially be used to rule out these theories~\cite{Barausse:2016eii}. We are extrapolating point-particle results to BH spacetimes. Such procedure was shown to be robust in GR
when the interacting objects are both BHs~\cite{Berti:2010ce,Tiec:2014lba,Sperhake:2011ik}. When stars are involved, interference effects decrease the total energy output~\cite{Haugan:1982fb,1985ApJS...58..297P}.

%%%%%%%%%%%%%%%%%%%%%%%%%%%%%%%%%%%%%%%%%%%%%%%%%
\section{Discussion} \label{sec:discussion}
%%%%%%%%%%%%%%%%%%%%%%%%%%%%%%%%%%%%%%%%%%%%%%%%%
We have worked out the details of gravitational radiation in theories of massive gravity,
when two BHs merge. It is clear that a substantial fraction of the radiation emitted in this process decays slowly, at large distances.
In fact, because the graviton is massive, low-energy GWs are confined. This was also noticed in Ref.~\cite{Sperhake:2017itk}.
Such radiation will clearly have an impact in any star or object located within its sphere of influence, but such effects are unknown to us so far. Circular motion at an orbital frequency $\omega=\mu$ will likely lead to resonant excitations of dipolar GWs.
Unfortunately, the numerical study of such resonances is a challenging task~\cite{Cardoso:2011xi,Yunes:2011aa,Fujita:2016yav}, upon which we did not embark.

Technically, our procedure is free of computational challenges. The perturbative framework that we use is an expansion in mass ratio. All the observables that we extract are finite, and tend to zero when the mass ratio decreases. Thus, perturbation theory is applicable and never breaks down as long as mass ratios are sufficiently small (in a well defined manner). The numerical results are converging and very clear: we show that new modes are excited to a substantial amplitude,
both in head-on collisions and in quasi-circular motion. For head-on collisions -- because new modes are excited at characteristically small frequencies -- GW detectors sensitive to low frequency radiation will be able to impose constraints on the mass of gravitons tighter than ever before. In fact, if we trust that our results carry over to two, nearly equal-mass neutron stars, then the constraints on the mass of the graviton will be improved by two orders of magnitude or more. The dipolar mode excited by quasi-circular inspirals is in fact dominant with respect to the GR quadrupolar emission. Thus, accurate observations of binary BHs have the potential to tightly constraint massive gravity.

Alternatively, our results can be a manifestation that the background geometry does not describe astrophysical BHs. Indeed, it can be shown that Schwarzschild (and Kerr) BHs are unstable in theories with a massive graviton~\cite{Babichev:2013una,Brito:2013wya,Brito:2013xaa}. Nevertheless, for small mass coupling $M\mu$ the instability timescale is extremely large and the spacetime responds to short-timescale phenomena ``unaware'' of the instability. Thus, sufficiently short-scale phenomena are expected to produce Schwarzschild BHs, and our methods and results apply in the regime where we would like them to, which is that of small graviton masses. 
In addition, numerical results suggest that when one of the metrics is taken to be non-dynamical, hairy stationary BHs do not even exist~\cite{Brito:2013xaa,Volkov:2016ehx}. One cannot exclude the possibility that a viable astrophysical BH is described by a dynamical metric~\cite{Rosen:2017dvn}, in which case our results could change considerably. In particular, Vainshtein screening -- absent in our background solutions -- may play a critical role in more generic background BH solutions. Notwithstanding, it is clear that GW astronomy carries a huge potential to understand theories of massive gravity: the existence of extra degrees of freedom lead in general to substantially different dynamics and gravitational-wave emission. To fully realize this potential several challenges (including the correct description of astrophysical BHs) need to be seriously tackled.

%%%%%%%%%%%%%%%%%%%%%%%%%%%%%%%%%%%%%%%%%%%%%%%%%
\begin{acknowledgments}
%%%%%%%%%%%%%%%%%%%%%%%%%%%%%%%%%%%%%%%%%%%%%%%%%
We are indebted to Evgeny Babichev, Claudia de Rham, Chris Moore, Andrew Tolley and Luis Lehner for many and very useful comments, 
discussions and suggestions. 
The authors acknowledge financial support provided under the European Union's H2020 ERC 
Consolidator Grant ``Matter and strong-field gravity: New frontiers in Einstein's theory'' grant 
agreement no. MaGRaTh--646597. 
This project has received funding from the European Union's Horizon 2020 research and innovation programme under the Marie Sklodowska-Curie grant agreement No 690904.
We acknowledge financial support provided by FCT/Portugal through grant PTDC/MAT-APL/30043/2017.
We acknowledge the SDSC Comet and TACC Stampede2 clusters through NSF-XSEDE Award Nos. PHY-090003.
The authors would like to acknowledge networking support by the GWverse COST Action CA16104, ``Black holes, gravitational waves and fundamental physics.''
\end{acknowledgments}

\bibliography{References}

\begin{thebibliography}{30}
\bibitem [{ext()}]{extraction_radii}%
  \BibitemOpen
  \href@noop {} {}\bibinfo {note} {We checked that the results scale very
  simply with extraction radius for equal to or larger than these. In fact, we
  made careful comparisons with the scaling for a simple massive scalar field.
  Our results are consistent and agree also with the results of
  Ref~\cite{Sperhake:2017itk}. We believe we are in the asymptotic regime where
  nonlinearities are not relevant.}\BibitemShut {Stop}%
\end{thebibliography}

\end{document}